\documentclass[10pt]{IEEEtran}

\ifCLASSINFOpdf

\fi

\usepackage[subrefformat=parens,labelformat=parens,caption=false,font=footnotesize]{subfig}
\usepackage{array}
\usepackage{balance}
\usepackage{amsmath}
\usepackage{breqn}
\usepackage[nolist,nohyperlinks]{acronym}
\usepackage{graphicx,wrapfig,lipsum}
\usepackage{rotating}
\usepackage{amsmath, amssymb, amsthm}
\usepackage{stmaryrd}
\usepackage{tikz}
\usepackage{hyperref}
\usepackage{verbatim}
\usepackage{url}
\usepackage[utf8]{inputenc}
\usepackage[english]{babel}

\usepackage{booktabs}

\usepackage{multicol}
\usepackage{xr-hyper}

\usepackage{scalerel}
\usepackage{multicol}

\usepackage[noadjust]{cite}
\usepackage[normalem]{ulem}
\usepackage{color}
\usepackage{dsfont}
\usepackage{bm}
\usepackage[short]{optidef} 
\usepackage{diagbox}
\usepackage{enumitem}
\usepackage{slashbox}
\usepackage[ruled]{algorithm2e}
\usepackage{multirow}
\usepackage[T1]{}
\usepackage{color, colortbl}
\usepackage{babel}
\usepackage{verbatim}
\usepackage{textcomp}
\usepackage{tabulary}
\usepackage{booktabs}
\usepackage{caption}
\definecolor{Gray}{gray}{0.95}
\definecolor{LightCyan}{rgb}{0.8,0.85,1}
\definecolor{LightBlue}{rgb}{0.6,0.6,1}
\usepackage[font=small]{caption}
\usepackage{mathtools}

\definecolor{GPT35context}{HTML}{F1B814}
\definecolor{GPT4}{HTML}{BD1E51}
\definecolor{GPT35}{HTML}{80ADCC}
\definecolor{constructCluster}{HTML}{2B83BA}

\definecolor{niceblue}{HTML}{007ED6}

\setlist{nosep}
\newcommand\blfootnote[1]{%
  \begingroup
  \renewcommand\thefootnote{}\footnote{#1}%
  \addtocounter{footnote}{-1}%
  \endgroup
}
\usepackage{mathtools}

\makeatletter
\def\blfootnote{\gdef\@thefnmark{}\@footnotetext}
\makeatother
\begin{document}

\title{TeleQnA: A Benchmark Dataset to Assess Large Language Models Telecommunications Knowledge}
\author{
Ali Maatouk$^{*\dagger}$, Fadhel Ayed$^{*\dagger}$, Nicola Piovesan$^\dagger$, Antonio De Domenico$^\dagger$, Merouane Debbah$^\ddagger$, Zhi-Quan Luo$^\mathsection$\\
$^\dagger$Paris Research Center, Huawei Technologies, Boulogne-Billancourt, France
\\ $^\ddagger$Khalifa University of Science and Technology, Abu Dhabi, UAE
\\ $^\mathsection$The Chinese University of Hong Kong, Shenzhen, China\vspace{-11pt}}
\maketitle
\thispagestyle{empty}
\blfootnote{$^*$Equal contribution.}
\begin{abstract}We introduce TeleQnA\footnote{Samples from this dataset can be found in Appendix \ref{appendix:examples}.}, the first benchmark dataset
designed to evaluate the knowledge of \acp{LLM} in telecommunications. Comprising 10,000 questions and answers, this dataset draws from diverse sources, including standards and research articles. This paper outlines the automated question generation framework responsible for creating this dataset, along with how human input was integrated at various stages to ensure the quality of the questions.
Afterwards, using the provided dataset, an evaluation is conducted to assess the capabilities of \acp{LLM}, including GPT-3.5 and GPT-4. The results highlight that these models struggle with complex standards-related questions but exhibit proficiency in addressing general telecom-related inquiries. Additionally, our results showcase how incorporating telecom knowledge context significantly enhances their performance, thus shedding light on the need for a specialized telecom foundation model. Finally, the dataset is shared with active telecom professionals, whose performance is subsequently benchmarked against that of the LLMs. The findings illustrate that \acp{LLM} can rival the performance of active professionals in telecom knowledge, thanks to their capacity to process vast amounts of information, underscoring the potential of \acp{LLM} within this domain. The dataset has been made publicly accessible on GitHub\footnote{https://github.com/netop-team/TeleQnA}.




\end{abstract}

\section{Introduction}
\label{sec:intro}
\acp{LLM} have recently sparked a revolution in the realm of \ac{NLP}, elevating automatic text generation and interaction to high levels of performance. Currently, the advent of \acp{LLM} has begun to impact many other fields beyond \ac{NLP}, such as medicine \cite{Singhal2023} and finance \cite{wu2023bloomberggpt}. Much like these fields, the telecom industry stands on the brink of experiencing the potential impact \acp{LLM} could have on its landscape. Indeed, it is foreseen that \acp{LLM} may greatly facilitate a wide range of tasks, including troubleshooting anomalies, comprehending standards, and providing optimization recommendations for network enhancement \cite{maatouk2023large,AltmanSolon2023,bariah2023understanding}.

The success of \acp{LLM} depends critically on benchmark datasets designed to assess their proficiency in specific domains. These datasets also play a pivotal role in determining the optimal architectural design for \acp{LLM} and guiding the pre-training procedure in these specialized fields. 
In the context of \ac{NLP} applications, notable examples include the TriviaQA \cite{Singhal2023}, HelloSwag \cite{zellers2019hellaswag}, and SIQA \cite{socialiqa} datasets, tailored to evaluate \acp{LLM} capabilities in reading comprehension, commonsense reasoning, and social intelligence, respectively. The same can be observed in other domains, such as medicine and finance, where benchmark datasets like MultiMedQA \cite{Singhal2023} and FLUE \cite{shah-etal-2022-flue} have been introduced to assess the proficiency of \acp{LLM} in these fields.

As \acp{LLM} find their way into the telecommunications industry, a clear and pressing issue arises—there is a notable absence of a benchmark dataset designed to evaluate these models' proficiency in telecom. Consequently, there is an urgent need for such a dataset, as highlighted in various prior research (e.g., \cite{kotaru2023adapting}). This paper aims to bridge this gap by introducing TeleQnA, the first benchmark dataset tailored specifically for evaluating \acp{LLM}' telecom knowledge. The contributions of our paper are fourfold:
\begin{itemize}
    \item First, we curate a comprehensive dataset that encompasses a vast corpus, spanning various disciplines within the telecommunications domain and taking on diverse forms, such as standards and research articles.
    \item Second, we introduce an automated \ac{QnA} generation process based on the gathered material, involving two \acp{LLM} communicating with one another. We show how this architecture, complemented by human input at different stages, not only ensures the quality of the generated dataset but also facilitates its scalability to accommodate a wide array of multidisciplinary topics.
    \item Third, we proceed to assess the telecom knowledge of prominent \acp{LLM}: GPT-3.5 and GPT-4 \cite{openai2023gpt4}. We illustrate their proficiency in responding to general telecom inquiries and highlight their deficiencies in addressing intricate questions related to standards specifications. Additionally, we demonstrate how the incorporation of contextual information significantly enhances these models' performance, particularly in areas where they typically face difficulties. All of these findings underscore the necessity for a specialized telecom foundation model to fully harness the potential of LLMs within the industry. 
    \item Lastly, we assess the performance of professionals actively engaged in this field in comparison to that of \acp{LLM}. Our findings demonstrate that \acp{LLM} can compete with active professionals in telecom knowledge. This potential stems from their ability to process extensive volumes of information, highlighting the valuable role \acp{LLM} are poised to play in shaping the future of this domain.
\end{itemize}
\section{Dataset Sources}
\label{sec:sources}
\begin{figure}
    \centering
\includegraphics[width=0.35\textwidth]{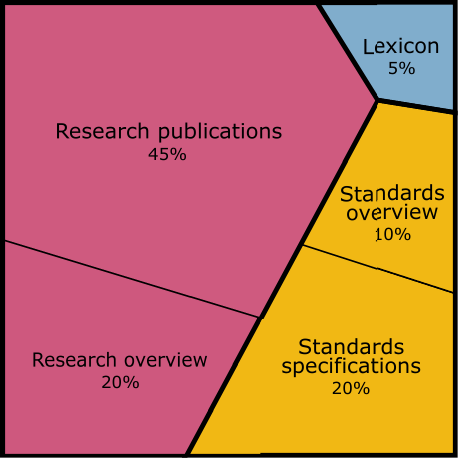}
    \caption{Distribution of the \ac{QnA} dataset among the categories of the collected source materials.}
    \label{fig:datasetdescription}
\end{figure}

When curating the dataset sources, our primary objective was to assemble a diverse and comprehensive collection of open-access data related to the telecom industry. This involved a broad spectrum of content, encompassing standards and scholarly research materials. In total, our collection comprised around 25,000 pages, containing approximately 6 million words. This approach guarantees that the resultant \ac{QnA} dataset offers an in-depth and well-rounded depiction of the telecom industry's multifaceted aspects. Furthermore, this diversity serves as a valuable resource for evaluating the strengths of an LLM in various aspects, including standard specifications and general telecom expertise. The allocation of the QnA dataset among the three specified categories – research, standards, and lexicon – is illustrated in Fig.~\ref{fig:datasetdescription}.

\subsection{Standards Documents}
Standards play a pivotal role in the field of telecommunications by ensuring that different technologies from various vendors can work together cohesively. These standards are established and maintained by recognized standardization bodies, including but not limited to 3GPP, IEEE, and ITU. Incorporating these documents when creating the \ac{QnA} dataset is immensely valuable for capturing the intricacies of the telecom industry. Particularly, it allows delving deep into the technical facets and the present-day implementation of telecom technologies. With this in mind, we have integrated two distinct document sets into our raw data, all closely associated with standard activities.
\begin{itemize}
    \item \textbf{Technical Specifications:} Technical specifications are detailed documents that define the technical standards for telecommunications systems. Given the sheer number of these documents, we have uniformly sampled a portion of them to ensure a non-biased representative selection of these documents across multiple standardization bodies.
    \item \textbf{Standards Overview:} To offer a more comprehensive perspective on standards, moving beyond the intricate technical specifications, we have gathered materials from various sources, encompassing standards summaries, review publications, and standards-related white papers.
\end{itemize}
\subsection{Research Material}
\begin{figure}
    \centering
    \includegraphics[width=0.50\textwidth]{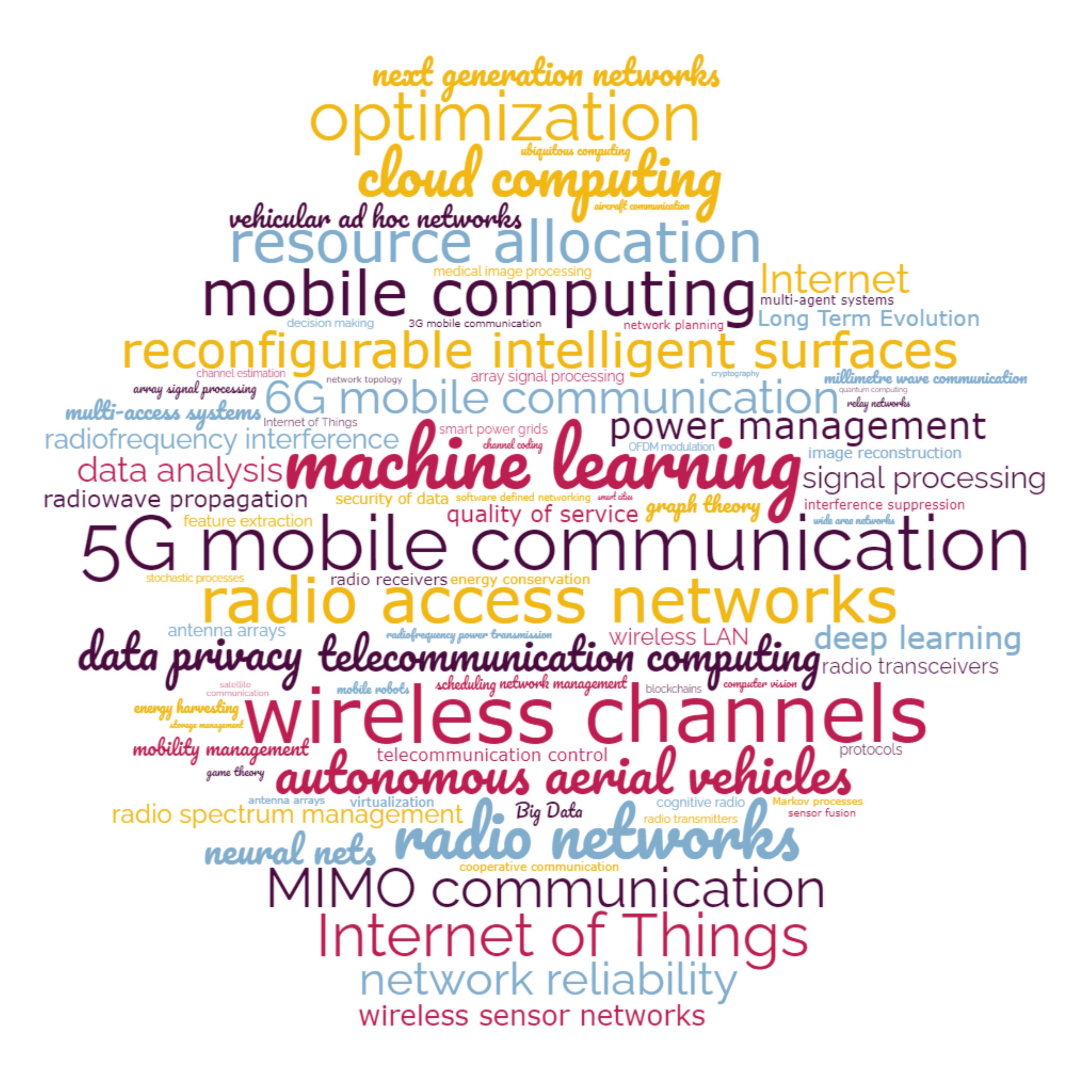}
    \caption{
    Word cloud illustrating the prominent keywords extracted from the collected research materials.}
    \label{fig:wordcloud}
\end{figure}
The landscape of publishing venues within the telecommunications field is vast. It encompasses a wide array of publishers, and is presented in diverse formats such as journals and conferences, with each repository housing a plethora of valuable information. The sheer abundance and diversity of material available pose a considerable challenge when attempting to create a \ac{QnA} dataset based on these scholarly works. Additionally, the varying levels of citation and peer review standards, coupled with our commitment to exclusively use open-access material, introduce another layer of complexity to the task of curating a reliable and representative dataset. To address these challenges, we have categorized our sources into two groups: \emph{Research publications} and \emph{Research overview}. The former category primarily comprises technical details sourced from research articles and in-depth technical open-access books, while the second includes content from review and survey venues. Within each of these categories, we have adhered to the subsequent set of guiding principles:
\begin{itemize}
    \item \textbf{Diversity:} Our first guiding principle was the diversity of material. In particular, we considered a broad spectrum of outlets, encompassing well-established 
    venues hosted by various publishers.
    \item \textbf{Relevance:} Within these outlets,
    we have chosen to identify the most influential material by considering their citation counts over the past 15 years. Our selection of this time frame is deliberate, as it allows us to incorporate the most recent and pertinent findings. On the other hand, the number of citations serves as an indicator of a manuscript popularity and impact over time. Furthermore, to infuse the dataset with the insights of industry experts, we have added material that telecom professionals have identified as pivotal contributions to this collection.
    \item \textbf{Bias Minimization:} To mitigate potential biases within the dataset, we have also taken a deliberate approach to uniformly sample a number of manuscripts from these venues. This strategy enhances the dataset diversity, encompassing not only popular topics but also those that, while less prominent, remain highly pertinent to the field.
    \item \textbf{Multi-disciplinarity:} We have ensured that our sources extend beyond the realm of pure communications and encompass topics relevant to the field. This expanded scope includes material from various domains, such as machine learning and optimization. This approach stems from the recognition that an \ac{LLM} tailored for the telecom sector should possess a degree of proficiency in interconnected domains. Although the majority of the sources consist of communication-oriented material, this diverse and multidisciplinary knowledge enriches the dataset, fortifying its comprehensive and interdisciplinary nature. 
\end{itemize}
A word cloud representing the keywords extracted from the considered research material is illustrated in Fig. \ref{fig:wordcloud}.
\subsection{Telecom Lexicon}
The telecommunications industry possesses a distinctive lexicon, characterized by an extensive array of technical terms, each holding significance in understanding the intricacies of communication networks. The incorporation of this lexicon into the generation of the \ac{QnA} dataset holds significant value, as it provides a perfect avenue to evaluate the proficiency of an \ac{LLM} in comprehending telecommunications terminology. To procure this source of data, we conducted a thorough examination of the standards documents and research papers we have collected and created a comprehensive dictionary of the technical terminology used therein. 
\section{Dataset Characteristics}
\label{sec:characteristics}
 \subsection{Question Type}
With the source material prepared, our first course of action involved adopting a question type, and after careful consideration, we opted for a multiple-choice approach. This decision was made for several reasons:
\begin{itemize}
   \item \textbf{Precision Assessment:} The multiple-choice format offers a straightforward means of gauging the accuracy of an \ac{LLM} when presented with the dataset. 
\item \textbf{Complex Decision-Making:} The multiple-choice format allows us to probe deeper into the LLM's decision-making capabilities. It enables us to assess whether the LLM can accurately discern situations where multiple choices may be correct, rather than just one.
\item \textbf{Nuanced Discrimination:} By adopting a multiple-choice framework, we can examine the LLM's ability to select the correct answer when presented with options that resemble the correct choice but are ultimately incorrect. 
\end{itemize}
For the reasons mentioned above, multiple-choice approaches have gained popularity in the realm of machine learning benchmarks. For instance, they have been prominently utilized in datasets like MCTest and SWAG \cite{richardson2013mctest,zellers2018swag}.
\subsection{Dataset Format}
To maintain consistency across all questions, we have implemented a standardized format. Each question is represented in JSON format, comprising five distinct fields:
\begin{itemize}
    \item \textbf{Question:} This field consists of a string that presents the question associated with a specific concept within the telecommunications domain.
    \item \textbf{Options:} This field comprises a set of strings representing the various answer options. 
    \item \textbf{Answer:} This field contains a string that adheres to the format ``option ID: Answer" and presents the correct response to the question. A single option is correct; however, options may include choices like ``All of the Above" or ``Both options 1 and 2". 
    \item \textbf{Explanation:} This field encompasses a string that clarifies the reasoning behind the correct answer.
    \item \textbf{Category:} This field includes a label identifying the source category depicted in Fig.
\ref{fig:datasetdescription}. 
\end{itemize}

\section{Dataset Creation}
\label{sec:creation}
\begin{figure*}
    \centering
    \includegraphics[width=0.8\textwidth]{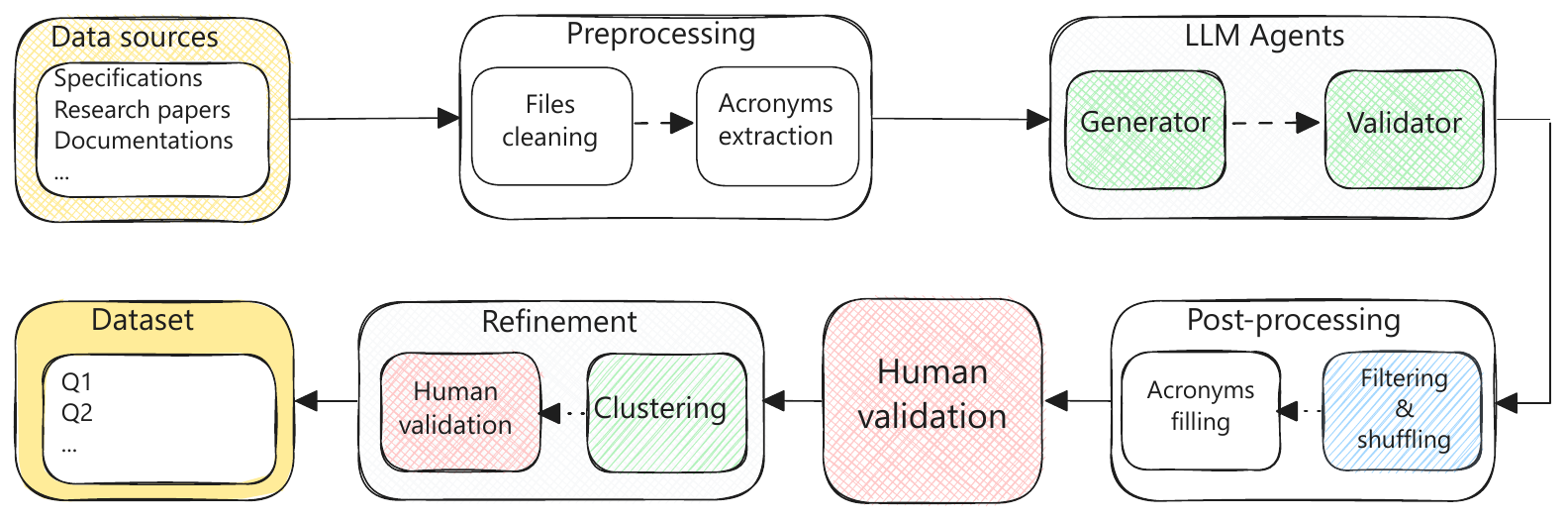}
    \caption{
    A high-level overview of the \ac{QnA} generation process.
    }
    \label{fig:generationprocess}

\end{figure*}

To construct a comprehensive \ac{QnA} dataset covering the multifaceted domain of telecommunications, a substantial number of questions is required. This task is further complicated by the specialized nature of telecom knowledge, demanding expertise to craft pertinent questions, answers, and explanations. Moreover, the telecom documents we collected often contain highly intricate information, making it infeasible for a team of human experts to generate questions and answers that comprehensively cover the diverse range of telecom subdomains. Adding to the complexity, we require these questions to be challenging to answer, and asking human experts to craft incorrect options that are also challenging across a wide array of subdomains is a task far from trivial. Given these challenges, adopting a crowdsourcing approach, similar to the one employed for the MCTest dataset \cite{richardson2013mctest}, becomes impractical.
To address these challenges, we leveraged OpenAI's GPT-3.5's API\footnote{The quality of the generation process was comparable between GPT-3.5 and GPT-4, which led us to favor the latter due to its lower cost.} 
to facilitate the generation process. This involved  developing an automated framework based on LLMs, which will be elaborated on in this section. An overview of the entire process can be found in Figure \ref{fig:generationprocess}.

\subsection{Preprocessing} 
Our initial step involved cleaning and formatting the collected data sources for seamless integration with the utilized LLMs. In parallel, we created acronyms extractors that identify the acronyms found in these documents and their corresponding definitions,  allowing us to grasp the terminology employed within these materials.
\subsection{\ac{LLM} Agents} 

Our LLM-based framework revolves around two GPT-3.5 LLM agents: a generator and a validator, each with distinct roles as outlined below. 
\begin{itemize}
    \item \textbf{Generator:} The generator is an \ac{LLM} instance responsible for producing multiple-choice questions. To accomplish this, it is supplied with a segment of a telecommunications document, selected from those discussed in Section \ref{sec:sources}. Subsequently, it is instructed to generate a set of multiple-choice questions based on the given context, adhering to the formatting guidelines. Additionally, it is directed to ensure that the questions are self-contained. Furthermore, it is prompted to provide a rationale for the correct answer, typically by referencing sentences within the document to substantiate the choice of the answer.
    \item \textbf{Validator:} The questions and options generated by the generator, together with the contextual information (i.e., the relevant segment of the document), are given as input to another GPT-3.5 \ac{LLM}, referred to as the validator. To prevent any biasing, the answer and the rationale behind the correct answer are intentionally omitted when providing input to the validator. The validator's primary function is to select an option among those provided solely based on its knowledge and the contextual information provided. If the validator picks the option designated as correct by the generator, the question is retained. However, if the validator selects an incorrect option, the question is discarded. This automated layer serves to enhance the overall accuracy and mitigate any potential instances of misinformation or fabrication that the generator may introduce. 
\end{itemize}
\subsection{Post-Processing}
Moving forward in the process, we transition into the post-processing stage. Following the initial validation step, this phase focuses on the following key aspects: 
\begin{itemize}
 \item \textbf{Filtering:} We developed a tool designed to identify and filter out questions that are not self-contained, meaning they refer to the text from which are taken from by including particular terms such as ``figure" and ``fig".
  \item \textbf{Shuffling:} Our experiments have revealed a tendency of GPT-3.5 to place the correct answer in option 1. To mitigate this bias, we shuffled the options to ensure equal likelihood for all choices.
    \item \textbf{Acronyms:} We designed acronym detectors that identify the use of acronyms in the questions and map them to their respective definitions. 
   
\end{itemize}
%

\subsection{Human Validation - First Stage} 
Having validated the questions using the validator and completed the necessary post-processing steps, we now introduce the first stage that incorporates human intervention. Specifically, a telecom expert is assigned the task of carefully reviewing each question, the provided answer options, the explanation given by the question generator for the answer, and the contextual information used. The objective is to make an informed decision regarding whether or not to discard a question. The primary objectives of this human-in-the-loop process are twofold:
\begin{itemize}
    \item \textbf{Ensure question correctness:} 
    The expert's role is to verify the accuracy of the questions, their associated options, and the designated correct answer, effectively eliminating improperly generated questions.
    \item \textbf{Ensure that the questions are self-contained:} The questions should not revolve around content that is of localized interest, such as experimental results specific to a proposed algorithm.
\end{itemize}

\subsection{Refinement}
The collected telecom sources inherently exhibit overlaps across the various venues. This overlap gives rise to redundancy within the \ac{QnA} dataset, a challenge that necessitates a refinement of the dataset. To address this issue systematically, we employed Open AI's Ada v2 text embeddings model to compute embeddings for all the questions in the \ac{QnA} dataset along with their corresponding explanations. These embeddings serve as compact numerical representations, encapsulating the essence of phrases, thus enabling us to gauge the semantic similarity between any two questions by measuring the proximity of their respective embeddings in this high dimensional space. Having computed these embeddings,  we then applied the K-Means clustering algorithm 
to group questions into clusters, in order to streamline the subsequent human intervention step. In fact, the concluding phase encompasses a secondary human validation that serves a dual purpose:
\begin{itemize}
    \item \textbf{Eliminating redundant questions:} The primary goal of the intervention is to eliminate duplicate or semantically-redundant questions within each cluster.
    \item \textbf{Final iteration of quality checking:} The second objective entails conducting a final round of verification, similar to the first human validation stage. 
\end{itemize}
\section{Performance Evaluation}
\begin{table}[t]
\centering
\begin{tabular}[t]{lcccccc}
\toprule
&Run 1 & Run 2 & Run 3 & Run 4 & Mean & Std\\
\midrule
50 questions & 63.35&	65.63&	65.10	&63.79		&64.46	&0.92 \\
25 questions &65.19	&67.56&	66.90	&66.30	&	66.48	&0.87 \\
10 questions &67.64&	64.32	&66.94	&67.20	&	66.52	&1.29 \\
5 questions &65.45	&67.55&	68.52&	66.85	&	67.09&	1.11 \\
1 question &66.43&	67.48&	68.10	&66.46	&	67.11	&0.70 \\
\bottomrule
\end{tabular}
\caption{Illustration of GPT-3.5's accuracy (\%) across multiple runs and question batch sizes.}
\label{table:variance}
\end{table}
\label{sec:performanceeval}
In this section, we conduct a comprehensive evaluation of GPT-3.5 and GPT-4 using the TeleQnA dataset. Our objective is to assess their proficiency in telecom knowledge while also offering insights into their strengths and weaknesses. Additionally, we compare the performance of telecom professionals to these language models, establishing a benchmark for reference. As a performance measure, we define the accuracy as the percentage of questions for which the entity at hand selected the option marked as correct in the dataset.
\subsection{Batch Size and Variance Study} 
Our numerical analysis begins by examining the variance of the accuracy achieved by the \acp{LLM} when responding to the TeleQnA dataset. Specifically, we are interested in tracking how the accuracy evolves with each iteration of \ac{LLM} questioning. Additionally, we investigate the impact of sending questions in random batches of varying sizes ($B=1,5,10,25,$ and $50$) to the \ac{LLM}. The former investigation aims to show the number of iterations required to obtain a statistically reliable accuracy score for the \ac{LLM}'s telecom knowledge. Meanwhile, the latter investigation seeks to identify trends when querying the \ac{LLM}'s API with larger question batches, potentially streamlining the process by sending questions in batches rather than one by one. To obtain these findings, we uniformly sampled 1000 questions and conducted accuracy assessments across multiple runs and with different batch sizes. The results of these experiments are presented in Table \ref{table:variance} for GPT-3.5\footnote{GPT-4 exhibited similar trends, and therefore, we have opted to exclude the specific details.}.

The results demonstrate the consistency of accuracy across multiple runs, with a standard deviation of approximately 1\%. Given that the standard deviation decreases with the number of questions considered, it can be inferred that this deviation is even smaller when applied to the entire dataset. Therefore, we can conclude that conducting a single run is sufficiently reliable for assessing the proficiency of the \ac{LLM}. 
Conversely, we observe a decline in mean accuracy when questions are submitted in larger batches. This decrease may be attributed to the inherent diversity of questions stemming from various sub-domains within telecommunications.
When questions are batched together, the \ac{LLM} encounters a broader spectrum of topics, which may result in less specialized responses and, therefore, a potential drop in accuracy. However, we note that this decline in performance remains modest when transitioning from a batch size of $B=1$ to $B=5$. Thus, we have chosen to utilize this batch size in the rest of our experiments to reduce the number of queries needed for assessment.
\subsection{Performance Benchmarks}
\label{subsec:perfbenchmark}
\begin{figure}
    \centering
    \includegraphics[trim=120 0 150 0,clip,
    width=0.5\textwidth]{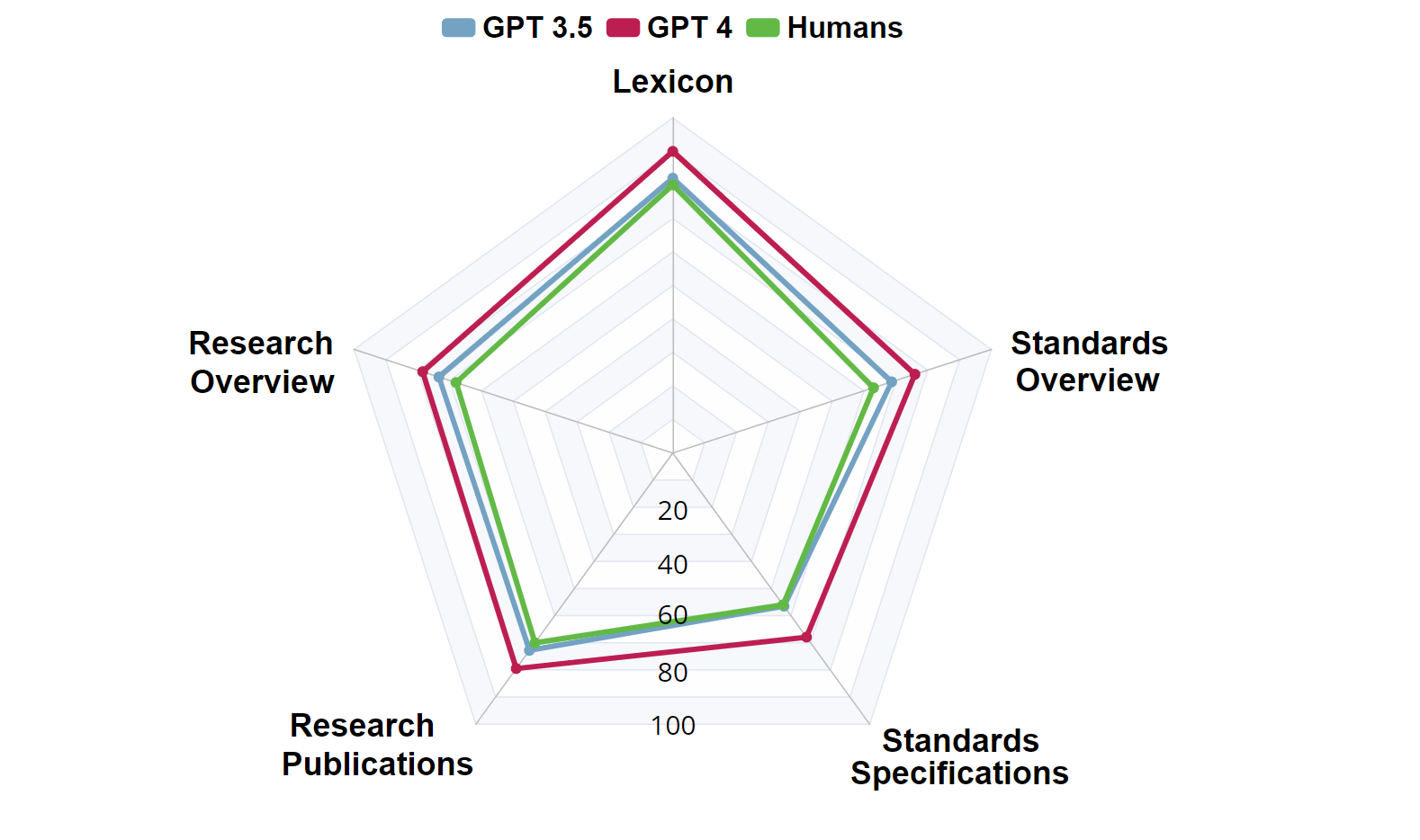}
    \caption{Accuracy (\%) comparison among GPT-3.5, GPT-4, and active professionals in all five categories.}
    \label{fig:bencharmking}
\end{figure}
We evaluate the performance of GPT-3.5 and GPT-4 across the various categories of the TeleQnA dataset. The results are reported in Fig. \ref{fig:bencharmking}, and the details can be found in Appendix \ref{appendix:benchmarksresults}. As anticipated, GPT-4 consistently outperforms GPT-3.5, demonstrating around 7\% improvement across all categories. Notably, \acp{LLM} exhibit exceptional performance in the lexicon category, which encompasses general telecom knowledge and terminology, achieving approximately 87\% accuracy for GPT-4. Conversely, these models face challenges when confronted with more intricate questions related to standards, with the highest performing model, GPT-4, achieving a modest 64\% accuracy in this domain. In summary, GPT-3.5 averaged an accuracy of 67\%, while GPT-4 achieved an accuracy of 74\%. These results demonstrate that these models possess a solid foundation in general telecom expertise. However, to attain higher accuracy in responding to complex inquiries, further adaptations to the telecom domain are necessary.

In our final step, we conducted a performance benchmark comparing active professionals to \acp{LLM}. To accomplish this, we designed surveys consisting of ten questions, with two questions representing each of the five categories. These surveys were then distributed to thirty active professionals in the telecom field, working in various domains such as standardization, signal processing, optical networks, etc. These professionals were explicitly instructed not to utilize search engines or external references, relying solely on their personal knowledge. The results reveal that \acp{LLM} and active professionals exhibit comparable performance in general telecom knowledge. However, in the case of intricate questions related to research and standards, \acp{LLM} demonstrate the capability to rival these professionals. This is attributed to \acp{LLM}' ability to digest and memorize complex and intricate documents. Furthermore, it is crucial to recognize the challenge faced by professionals when responding to these questions, as they encompass a broad range of telecom subdomains that these individuals may not be necessarily actively engaged with in their work. Considering all factors, our results underscore the significant promise that \acp{LLM} hold within this domain, as demonstrated by their competitiveness within this extensive and comprehensive dataset.
\subsection{Influence of Context}
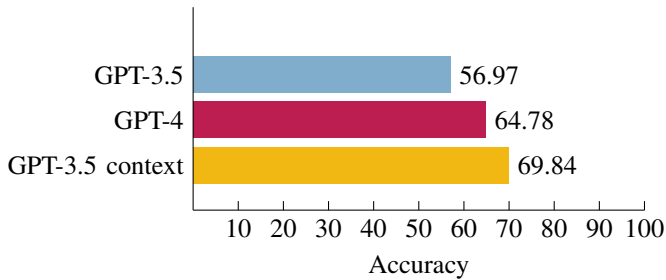
\begin{figure}
    \centering
    \begin{tikzpicture}[x={(.1,0)},scale=0.6]
\foreach  \l/\x/\c[count=\y] in {GPT-3.5 context/69.84/GPT35context, 
GPT-4/64.78/GPT4, 
GPT-3.5/56.97/GPT35}
{\node[left] at (0,\y) {\l};
\fill[\c] (0,\y-.4) rectangle (\x,\y+.4);
\node[right] at (\x, \y) {\x};}
\draw (0,0) -- (100,0);
\foreach \x in {10, 20, ..., 100}
{\draw (\x,.2) -- (\x,0) node[below] {\x};}
\draw (0,0) -- (0,4.5);
\node[below] at (50,-0.8) {Accuracy};
\end{tikzpicture}
    \caption{Accuracy (\%) comparison among GPT-3.5, GPT-4, and GPT-3.5 with context in the standards specifications category.}
    \label{fig:contextperformance}
\end{figure}
Until this stage of our benchmarking process, we have been querying the \acp{LLM} without accompanying contextual information for the questions. Nevertheless, in this subsection, our goal is to investigate how supplementing questions with additional context affects the accuracy of these models.

To accomplish this, we have focused on the standards specifications sources, encompassing thousands of technical standards pages. Our selection of this category is driven by the fact that this is where \acp{LLM} have exhibited the lowest performance. With this in mind, we segmented these pages into approximately 500-word segments before generating embeddings for each segment using OpenAI's Ada v2 text embeddings. Moreover, we employed the same OpenAI model to create embeddings for the questions and corresponding options belonging to this category. Following that, we constructed a distance matrix between the embeddings of each question-options pair and those of each segment. The next step involved querying the \ac{LLM} by supplying batches of five questions-options pairs, and additionally, as context, the top-3 closest segments to the five question-option pairs based on the distance matrix.
The outcomes of these experiments are illustrated in Fig. \ref{fig:contextperformance}.

The 22.5\% relative accuracy gain highlights the significant enhancement in performance achieved by incorporating contextual information, demonstrating how even less advanced models like GPT-3.5 can match the performance of state-of-the-art GPT-4 model. This underscores the necessity for a specialized telecom language models, fine-tuned or trained specifically on telecom-related data. Developing such a foundation model has the potential to push the boundaries of \acp{LLM} performance in the telecom domain, paving the way for a wide range of use cases that demand telecom knowledge and expertise.
\section{Conclusions, Limitations, and Future Directions}
\label{sec:conclusions}
\subsection{Conclusions}
In this paper, we introduced TeleQnA, the first open-source benchmark dataset tailored specifically to the telecom industry. With a collection of 10,000 questions and answers drawn from various sources in the field, TeleQnA serves as an evaluation tool for assessing the knowledge of \acp{LLM} in the telecommunications domain. Throughout this paper, we have detailed the development process of TeleQnA and conducted an extensive evaluation of GPT-3.5 and GPT-4 using TeleQnA. The results have revealed interesting insights: although these models excel in addressing general telecom-related questions, they face challenges when tackling complex inquiries. Importantly, our findings emphasize the critical need for specialized telecom foundation models.
\subsection{Limitations}
While our framework's design is aimed at minimizing errors in the questions, it remains dependent on human judgment to determine whether a generated question should be retained or discarded. This introduces an element of subjectivity, particularly in cases where questions offer multiple valid options, albeit with varying degrees of correctness. In such instances, human subjectivity comes into play. For instance, let us consider the following question:\\

\noindent \textbf{Question 1:} What advantages does a data center-enabled HAP (High Altitude Platform) offer from an energy perspective?
\begin{itemize}
    \item \emph{Option 1:} Saves cooling energy and harvests solar energy
    \item \emph{Option 2:} Uses renewable energy and offsets carbon emissions
    \item \emph{Option 3:} Deploys large-scale solar panels and batteries
    \item \emph{Option 4:} Requires less energy for communication links
\end{itemize}
\textbf{Answer:} \emph{Option 1:} Saves cooling energy and harvests solar energy\\
\textbf{Explanation:} The data center-enabled HAP saves cooling energy due to its location at the naturally low temperature stratosphere and harvests solar energy using large solar panels.\\
\textbf{Category:} Research publications
\\

An expert might discern that Option 1 is more accurate than Option 2. In fact, although Option 2 is indeed an advantage offered by data center-enabled HAP, Option 1 provides more insights into the fact that high-altitude platforms are deployed in the low temperature stratosphere, while also considering the fact that solar energy is used (i.e., a renewable energy). We eliminated some of these questions to ensure the utmost objectivity and correctness, but deliberately retained others to challenge the \ac{LLM} and assess its capability to discern between options exhibiting differing degrees of correctness.

Another important consideration is that a substantial portion of the generated questions draw from research papers to encompass a wide range of subjects. While this enriches the dataset, it introduces an element of subjectivity, as authors in their papers may present their own perspectives and interpretations of phenomena, which can be inherently subjective. While our aim has been to filter out such questions, it is important to acknowledge that, in the end, the process involves an element of human subjectivity. All of these factors collectively indicate that, despite our efforts to optimize the process extensively, we must acknowledge that the dataset generation is not flawless, as it continues to rely on human input.
\subsection{Future Directions}
In light of these limitations, one thing becomes abundantly clear: the development and advancement of extensive and comprehensive telecom knowledge datasets for evaluating \acp{LLM} necessitate a collaborative effort spanning the entire industry. We are receptive to all forms of feedback, especially in cases where certain questions may elicit nuanced perspectives from active community members regarding their correctness. In the end, we view TeleQnA as an initial step toward such a collaborative endeavor. Researchers also have the potential to create their own datasets, and by combining these diverse datasets with TeleQnA, one can see a future where an expansive and all-encompassing dataset, akin to those successfully crafted by the machine learning community for their applications of interest, can be created. This, in turn, paves the way for the widespread adoption of high-performing \acp{LLM} tailored specifically for telecom applications. 
\section{Acknowledgments}
We express our gratitude to the active telecom professionals for their generous contribution of time and participation in our surveys designed to evaluate telecom professionals' knowledge using TeleQnA.

\bibliographystyle{IEEEtran}

\bibliography{reference.bib}
\newpage
\vspace{0cm}
\begin{IEEEbiographynophoto}
    {Ali Maatouk} (Member, IEEE)
is a Researcher with Huawei Technologies, France. His research interests include the mathematical modeling and optimization of communication systems, machine learning, and the notion of information freshness.
\end{IEEEbiographynophoto}
\vspace{-1.1cm}
\begin{IEEEbiographynophoto}
    {Fadhel Ayed} is a Senior Researcher with Huawei Technologies, France. His research interests include resource-efficient machine learning, stochastic processes, and mathematical foundations of deep learning.
\end{IEEEbiographynophoto}
\vspace{-1.1cm}
\begin{IEEEbiographynophoto}
    {Nicola Piovesan} (Member, IEEE)
is a Senior Researcher with Huawei Technologies, France. His research interests include energy sustainability, energy efficiency optimization, and machine learning in wireless communication systems.
\end{IEEEbiographynophoto}
\vspace{-1.1cm}
\begin{IEEEbiographynophoto}
    {Antonio De Domenico} (Member, IEEE)
is a Senior Researcher with Huawei Technologies, France. His research interests include heterogeneous wireless networks, machine learning, and green communications. 
\end{IEEEbiographynophoto}
\vspace{-1.1cm}
\begin{IEEEbiographynophoto}
    {M\'erouane Debbah}
    (Fellow, IEEE) is a Professor at  Khalifa University of Science and Technology in Abu Dhabi. His research interests include Large Language Models, distributed AI systems for networks and semantic communications.
\end{IEEEbiographynophoto}

\begin{IEEEbiographynophoto}
    {Zhi-Quan Luo}
    (Fellow, IEEE) is is Vice President (Academic) of the Chinese University of Hong Kong (Shenzhen). His research interests lie in optimization algorithms for signal processing, wireless
communication and data analytics.
\end{IEEEbiographynophoto}

\begin{acronym}[AAAAAAAAA]
  \acro{3GPP}{Third Generation Partnership Project}
  \acro{QnA}{Question and Answer}
 \acro{AI} {Artificial Intelligence}
 \acro{BERT}{Bidirectional Encoder Representations from Transformers}
  \acro{DL}{Deep Learning}
   \acro{FPGA}{Field-Programmable Gate Array}
  \acro{GPT}{Generative Pre-trained Transformer}
  \acro{KPI}{Key Performance Indicator}
 \acro{LLM}{Large Language Model}
  \acro{MNO}{Mobile Network Operator}
\acro{BS}{Base Station}
  \acro{ML}{Machine Learning}
 \acro{NLP}{Natural Language Processing} 
 \acro{API}{Application Programming Interface} 
 \acro{RAN}{Radio Access Network}
 \acro{MIMO}{multiple-input multiple-output}

 \end{acronym}


\appendices
\section{Samples of TeleQnA}
\label{appendix:examples}
\noindent \textbf{Question 1:} What does EIRP stand for?
\begin{itemize}
    \item \emph{Option 1:} Effective Isotropic Radio Power
    \item \emph{Option 2:} Equivalent Isotropic Radiated Power
    \item \emph{Option 3:} 
    Efficient Isotropic Radiation Propagation
    \item \emph{Option 4:} Effective Infrared Radiation Power
    \item \emph{Option 5:} Equivalent Infrared Radiated Power
\end{itemize}
\textbf{Answer:} \emph{Option 2:} Equivalent Isotropic Radiated Power\\
\textbf{Explanation:} EIRP stands for Equivalent Isotropic Radiated Power.\\
\textbf{Category:} \color{niceblue} \textbf{Lexicon}\color{black}
\\

\noindent \textbf{Question 2:} What is a Heterogeneous Network?
\begin{itemize}
    \item \emph{Option 1:} A network consisting of multiple cells with different characteristics
    \item \emph{Option 2:} A network that provides services to the user in a managed way
    \item \emph{Option 3:} 
    A network that changes the radio access mode or system used for bearer services
    \item \emph{Option 4:} A network that broadcasts a specific indicator and identity
    \item \emph{Option 5:} A network where the subscriber’s personal service environment is controlled
\end{itemize}
\textbf{Answer:} \emph{Option 1:} A network consisting of multiple cells with different characteristics\\
\textbf{Explanation:} A heterogeneous network is a 3GPP access network with multiple cells having different characteristics.\\
\textbf{Category:} \color{niceblue} \textbf{Lexicon}\color{black}
\\

\noindent \textbf{Question 3:} Which communication technique uses Light Diodes as transmitters?
\begin{itemize}
    \item \emph{Option 1:} Large-Scale Antenna Arrays
    \item \emph{Option 2:} Free Space Optical Communications
    \item \emph{Option 3:} 
    Heterogeneous Networks
    \item \emph{Option 4:} Cooperative Relay Communications
    \item \emph{Option 5:} Cognitive Radio Communications
\end{itemize}
\textbf{Answer:} \emph{Option 2:} Free Space Optical Communications\\
\textbf{Explanation:} Free Space Optical Communications use Light Diodes as transmitters.\\
\textbf{Category:} \color{niceblue} \textbf{Research overview}\color{black}
\\

\noindent \textbf{Question 4:} What is the main advantage of using SDR (software-defined radio) techniques?
\begin{itemize}
    \item \emph{Option 1:} They provide higher computational power for signal processing.
    \item \emph{Option 2:} They are more cost-effective compared to traditional communication systems.
    \item \emph{Option 3:} 
    They offer flexibility to update receiver and transmitter functionalities by modifying software code.
    \item \emph{Option 4:} They result in lower latency in signal processing.
    \item \emph{Option 5:} None of the above.
\end{itemize}
\textbf{Answer:} \emph{Option 3:} They offer flexibility to update receiver and transmitter functionalities by modifying software code.\\
\textbf{Explanation:} SDR platforms are designed to be highly flexible, where all the receiver and transmitter functionalities can be updated by a simple modification of the software code.\\
\textbf{Category:} \color{niceblue} \textbf{Research overview}\color{black}
\\

\noindent \textbf{Question 5:} What is the maximum number of eigenmodes that the MIMO channel can support? (nt is the number of transmit antennas, nr is the number of receive antennas)
\begin{itemize}
    \item \emph{Option 1:} nt
    \item \emph{Option 2:} nr
    \item \emph{Option 3:} 
    min(nt, nr)
    \item \emph{Option 4:} max(nt, nr)
\end{itemize}
\textbf{Answer:} \emph{Option 3:} min(nt, nr)\\
\textbf{Explanation:} The maximum number of eigenmodes that the MIMO channel can support is min(nt, nr).\\
\textbf{Category:} \color{niceblue} \textbf{Research publications}\color{black}
\\

\noindent \textbf{Question 6:} What are the parameters governing the wireless channel propagation?
\begin{itemize}
    \item \emph{Option 1:} Positions of the transmitter and receiver antenna elements
    \item \emph{Option 2:} Carrier frequency
    \item \emph{Option 3:} 
    Nature and positions of scattering objects in the environment
    \item \emph{Option 4:} All of the above
\end{itemize}
\textbf{Answer:} \emph{Option 4:} All of the above\\
\textbf{Explanation:} The parameters governing the wireless channel propagation include the positions of the transmitter (Tx) and receiver (Rx) antenna elements, the carrier frequency, as well as the nature and positions of scattering objects in the environment.\\
\textbf{Category:} \color{niceblue} \textbf{Research publications}\color{black}
\\

\noindent \textbf{Question 7:} What frequency band does Bluetooth use? [Bluetooth - Overview]
\begin{itemize}
    \item \emph{Option 1:} 2.4 GHz
    \item \emph{Option 2:} 5 GHz
    \item \emph{Option 3:} 
    40 MHz
    \item \emph{Option 4:} 1 MHz
\end{itemize}
\textbf{Answer:} \emph{Option 1:} 2.4 GHz\\
\textbf{Explanation:} Bluetooth uses the 2.4 GHz licence-free ISM band for transmission.\\
\textbf{Category:} \color{niceblue} \textbf{Standards overview}\color{black}
\\

\noindent \textbf{Question 8:} Which pairs of wires are used in 10/100Base-T? [IEEE 802.3]
\begin{itemize}
    \item \emph{Option 1:} Pair 1 and pair 2
    \item \emph{Option 2:} Pair 2 and pair 3
    \item \emph{Option 3:} 
    Pair 3 and pair 4
    \item \emph{Option 4:} Pair 4 and pair 1
\end{itemize}
\textbf{Answer:} \emph{Option 2:} Pair 2 and pair 3\\
\textbf{Explanation:} In 10/100Base-T, pair 2 (orange/white and orange) and pair 3 (green/white and green) are used.\\
\textbf{Category:} \color{niceblue} \textbf{Standards overview}\color{black}
\\

\noindent \textbf{Question 9:} What is the RRC buffer size for a UE? [3GPP Release 17]
\begin{itemize}
    \item \emph{Option 1:} 45 MB
    \item \emph{Option 2:} 45 KB
    \item \emph{Option 3:} 
    45 GB
    \item \emph{Option 4:} 45 TB
    \item \emph{Option 4:} 4500 KB
\end{itemize}
\textbf{Answer:} \emph{Option 2:} 45 KB\\
\textbf{Explanation:} The RRC buffer size for a UE is 45 Kbytes.\\
\textbf{Category:} \color{niceblue} \textbf{Standards specifications}\color{black}
\\

\noindent \textbf{Question 10:} What are the required inputs for the Nadrf\_MLModelManagement\_Delete service operation? [3GPP Release 18]
\begin{itemize}
    \item \emph{Option 1:} Storage Transaction Identifier
    \item \emph{Option 2:} Unique ML Model identifier(s)
    \item \emph{Option 3:} 
    Model file address(es)
    \item \emph{Option 4:} Storage Transaction Identifier or Unique ML Model identifier(s)
    \item \emph{Option 4:} Storage Transaction Identifier and Unique ML Model identifier(s)
\end{itemize}
\textbf{Answer:} \emph{option 4:} Storage Transaction Identifier or Unique ML Model identifier(s)\\
\textbf{Explanation:} The required inputs for the Nadrf\_MLModelManagement\_Delete service operation are either the Storage Transaction Identifier or the Unique ML Model identifier(s).\\
\textbf{Category:} \color{niceblue} \textbf{Standards specifications}\color{black}
\section{Detailed Benchmarking Results}
\label{appendix:benchmarksresults}
In this appendix, we report the results obtained from our experiments involving GPT-3.5 and GPT-4 queries using TeleQnA. These results were obtained by randomly sampling batches of five questions drawn from TeleQnA, which were then provided to the large language model, and subsequently compared to the actual, correct answers.
\begin{table}[ht!]
\centering
\begin{tabular}[t]{lccc}
\toprule
& GPT-3.5 & GPT-4 & Humans \\
\midrule
\hspace{-5pt}Lexicon (500 questions) & 82.20 & 86.80  & 80.33 \\
\hspace{-5pt}Research overview (2000 questions) & 68.50 & 76.25  & 63.66 \\
\hspace{-5pt}Research publications (4500 questions) & 70.42 & 77.62 & 68.33 \\
\hspace{-5pt}Standards overview (1000 questions) & 64.00 & 74.40 & 61.66\\
\hspace{-5pt}Standards specifications (2000 questions) & 56.97 & 64.78 & 56.33 \\
\hspace{-5pt}Overall accuracy (10000 questions) & 67.29 & 74.91  & 64.86\\
\bottomrule
\end{tabular}
\caption{Illustration of GPT-3.5, GPT-4, and active professionals accuracy (\%) across the various TeleQnA categories.}
\label{table:detailed}
\end{table}
\end{document}